\documentclass[11pt,fleqn]{article}
\usepackage{amsfonts,amssymb,cite}
\usepackage{epsf,graphicx}

\def\noi{\noindent}

\newcommand{\Title}[1]{\noi {\uppercase{\Large #1}}\\[1ex]}

\newcommand{\Author}[2]{\noi{\large\bf #1}\\[2ex]\noi{\normalsize\it #2}\\}

\newcommand{\Rec}[1]{\noi {\it Received #1} \\}

\newcommand{\Abstract}[1]{\vskip 2mm \begin{center}
        \parbox{16.4cm}{\small\noi #1} \end{center}\medskip}
\newcommand{\PACS}[1]{\begin{center}{\small PACS: #1}\end{center}}
\newcommand{\foom}[1]{\protect\footnotemark[#1]}

\def\email#1#2{\footnotetext[#1]{e-mail: #2}\addtocounter{footnote}{1}}

\def\nqq{\hspace*{-2em}}
\def\nhq{\hspace*{-0.5em}}

\def\cm{\hspace*{1cm}}

\def\Acknow#1{\subsection*{Acknowledgment} #1}

\def\Jl#1#2{#1 {\bf #2},\ }

\def\ApJ#1 {\Jl{Astroph. J.}{#1}}
\def\CQG#1 {\Jl{Class. Quantum Grav.}{#1}}
\def\DAN#1 {\Jl{Dokl. AN SSSR}{#1}}
\def\GC#1 {\Jl{Grav. \& Cosmol.}{#1}}
\def\GRG#1 {\Jl{Gen. Rel. Grav.}{#1}}
\def\JETF#1 {\Jl{Zh. Eksp. Teor. Fiz.}{#1}}
\def\JETP#1 {\Jl{Sov. Phys. JETP}{#1}}
\def\JHEP#1 {\Jl{JHEP}{#1}}
\def\JMP#1 {\Jl{J. Math. Phys.}{#1}}
\def\NPB#1 {\Jl{Nucl. Phys.}{B\ #1}}
\def\NP#1 {\Jl{Nucl. Phys.}{#1}}
\def\PLA#1 {\Jl{Phys. Lett.}{#1A}}
\def\PLB#1 {\Jl{Phys. Lett.}{#1B}}
\def\PRD#1 {\Jl{Phys. Rev.}{D\ #1}}
\def\PRL#1 {\Jl{Phys. Rev. Lett.}{#1}}

\def\al{&\nhq}
\def\lal{&&\nqq {}}
\def\eq{Eq.\,}
\def\eqs{Eqs.\,}
\def\beq{\begin{equation}}
\def\eeq{\end{equation}}
\def\bear{\begin{eqnarray}}
\def\bearr{\begin{eqnarray} \lal}
\def\ear{\end{eqnarray}}
\def\earn{\nonumber \end{eqnarray}}

\def\nnn{\nonumber\\ \lal }

\def\yy{\\[5pt] {}}
\def\yyy{\\[5pt] \lal }
\def\eql{\al =\al}

\def\const{{\rm const}}

\begin{document}
\thispagestyle{empty}
\twocolumn[

\Title{On calculation of the cosmological parameters\yy
    in exact models of inflation.}

\Author{S. V. Chervon\foom 1 and I. V. Fomin\foom2}
    {Ulyanovsk, Ulyanovsk State University}

\Rec{October 31, 2007}

\Abstract {We discuss a method of calculating the key cosmological
  parameters on the basis of a selected scale factor by using exact
  solutions of the background equations. We specify the formulas for
  calculating the power spectrum, the spectral indices and their ratios,
  and the ratio of squared amplitudes of scalar and tensor perturbations.
  We obtain, for the first time, expressions for the basic cosmological
  parameters in terms of the scalar factor value at the instant of crossing
  the cosmological horizon. In this case, the cosmological parameters are
  calculated for a wide class of exact models. They are compared with their
  analogs derived in the slow-rolling approximation.

}

] %%%%%%%%%%%%%%%%%%%%%%%%%%%%%%%%%%%%%%%
\email 1 {sv chervon@rambler.ru}
\email 2 {ingvor@inbox.ru}

\section{Introduction.}

  The progress achieved in inflationary cosmology \cite{St80plus} enables
  us to compare the observational data with predictions of the theory:
  having chosen a cosmological model describing the Universe at its
  inflationary stage, one can determine the power spectrum of density
  perturbations, the spectral indices of scalar and tensor perturbations and
  their ratio as well as the ratio of squared amplitudes of the tensor
  and scalar modes. The basic theoretical conclusions have been obtained, as
  a rule, in the framework of the slow-roll approximation \cite{liddle93}.
  A method of obtaining new theoretical conclusions on the basis of
  exact solutions was proposed comparatively recently \cite{ch04grg}, and
  this method has enabled us to introduce corrections to calculations of
  the perturbation spectrum at horizon crossing. An exact expression for
  calculating the spectral index in terms of the total energy of the
  scalar field was found in \cite{chnotr05} for models of power-law and
  exponential-power-law inflation. It was shown \cite{ch07ir} how to obtain
  the key cosmological parameters at horizon crossing on the basis of a
  selected scale factor or, which is the same, the Hubble parameter, in the
  spirit of the potential fine tuning method \cite{ch97m,czs97plb} (see also
  \cite{ellis91,barrow94,maartens95}). The cosmological parameters were
  calculated and compared with their analogs obtained in the slow-rolling
  approximation for power-law and exponential-power-law inflation and for
  singular and nonsingular representations of de Sitter cosmology.

  In this paper, we widen the circle of the exact models studied and discuss
  the possibility of re-calculating the cosmological parameters at
  crossing the horizon for comparison with the observational data.

\section{The method of calculating the cosmological parameters}

  Initial perturbations of a cosmological model of inflation are described
  by a self-consistent set of the Einstein equations and those of a
  self-interacting scalar field in Friedman-Robertson-Walker space-time with
  flat spatial sections. Let us choose a presentation of this system
  in terms of the total energy potential $W(\phi)$ \cite{zhuche00j}:
\bear   \label{potW-2}
        W(\phi) \eql 3 M_P^2 H^2,
\\    \label{pot-3}
        3H\dot\phi \eql -\frac{d}{d\phi}W(\phi),
\\    \label{potW-2a}
    [\dot\phi(t)]^2 \eql -2M_P^2\dot H,
 \ear
  Recalled that the total energy potential has been defined as a function
  of the scalar field
\beq  \label{ConVV}
      W(\phi)=V(\phi)+\frac{1}{2}U^2(\phi), \cm U(\phi)=\dot\phi
\eeq

  It is easy to verify that one of the Einstein equations (3) may be
  obtained as a consequence of \eqs (1) and (2).

  Taking into account an exact formula for differential of the logarithm of
  the wave vector at horizon crossing time, $k=aH$, valid for perturbations
  with the wave vector $k$ \cite{chnotr05},
\bearr
    d(\ln k )= H dt + \frac{\dot{H}}{H}dt
            =Hdt+ \frac{3U^3(\phi)}{2M^2_P W'} dt,
\nnn
\ear
  one can make an amendment in the expression for the curvature
  perturbations power spectrum by substituting the physical potential
  $V(\phi)$ with the total energy potential $W(\phi)$; then, using \eqs
  (1)--(3), we can obtain a representation in terms of the Hubble parameter:
\beq
    {\mathcal{P}}_{\mathcal{R}}(k)= \frac{1}{12
      \pi^2M_P^6}\frac{W^3}{W'^2}=-\frac{H^4}{8M_{p}^2\dot{H}}\bigg|_{k=aH}
\eeq
  For the spectral index of the scalar perturbations, we obtain in a
  similar way
\bearr \label{ns}
    n_S(k)-1=  \frac{{d \ln \mathcal{P}}_{\mathcal{R}}(k)}{d\ln k}
\nnn
    = \frac{M_P^2 W'}{\frac{3}{2}U^2-W}
    \left(3\frac{W'}{W}-2\frac{W''}{W'} \right) = \frac{4\dot{H} -
        \frac{H\ddot{H}}{\dot{H}}}{\dot{H}+H^2}\bigg|_{k=aH}.
\nnn
\ear
  On the basis of these relations, one can evidently obtain expressions for
  the runaway of spectral indices.

  A procedure similar to that of calculating the power spectrum of scalar
  perturbations makes it possible to derive an exact expression for the
  power spectrum of tensor perturbations ${\mathcal{P}}_{\mathcal{G}}(k)$,
\beq
    {\mathcal{P}}_{\mathcal{G}}(k)=
    \frac{W}{6\pi^2M_P^4}=\frac{H^2}{2\pi^2 M_p^2}\bigg|_{k=aH},
\eeq
  and to write down a formula for the spectral index of tensor perturbations:
\beq\label{ng}
    n_G=M_P^2\frac{(W')^2}{W(\frac{3}{2}U^2-W)}
        = \frac{2\dot{H}}{\dot{H}+H^2}\bigg|_{k=aH}.
\eeq

  Let us note a specific result for the ratio of the spectral indices of
  tensor and scalar perturbations:
\beq
    r:=\frac{n_G}{n_S}=2\dot{H}\biggl[5\dot{H}
            +H^2-\frac{H\ddot{H}}{\dot{H}}\biggr]^{-1}.
\eeq
  Since the improvement in \eqs (\ref{ns}) and (\ref{ng}) consists in that
  the denominator contains $(\dot H + H^2)$ rather than $H^2$, the
  improvement does not manifest itself in this ratio.

  In what follows, we will consider the tensor-to-scalar ratio for
  squared amplitudes \cite{lukash06}
\beq
     \frac{T}{S}=\frac{A_T^2}{A_S^2} = -4\frac{\dot{H}}{H^2}
\eeq
  One can determine the amplitude of the tensor mode and estimate its
  contribution to the CMB anisotropy on the basis of the tensor-to-scalar
  ratio and the observed contribution of the scalar mode.

\section{Post-inflationary evolution of cosmological perturbations}

  Consider the sequence of events at transition from the inflationary stage
  to the radiation and matter domination stages.

  The inflationary stage is completed by scalar field decay and particle
  production, followed by nucleosynthesis and further evolution according to
  the standard scenario. Cosmological perturbations of different wavelengths
  (with different wavenumbers $k$) become classical quantities during
  several e-folds after the instant of coming outside the horizon. This time
  is designated as $t_*$.

  Having crossed the horizon, the cosmological perturbations remain
  ``frozen'' in the gravitational background and do not change their
  magnitude in the comoving reference frame. The magnitude of perturbations
  evolves in a known way in the radiation-dominated epoch at entering under
  the horizon, an we denote this time as $t_{\rm pr}$.

  The theory of cosmological perturbations is considered to be applicable
  in the initial epoch which begins before the cosmological scale of
  interest enters the horizon. This initial epoch begins much later than the
  nucleosynthesis, and therefore the matter constituents of the Universe,
  except nonbaryonic dark matter, are known. It has been established
  theoretically how perturbations of all components of the Universe evolve
  after the initial epoch evolve if the energy density of each component
  at that time is known.

  The initial perturbation spectrum can be constructed using the transfer
  function from vacuum fluctuations at the instant $t_*$. It is known how to
  calculate this function; in doing so, the perturbation of each component
  is calculated from the curvature perturbation $\mathcal{R}_{\bf k}$ with
  the aid of the transfer function:
\[
    g_{\bf k}(t)=T_g(t,k)\mathcal{R}_{\bf k},
\]
  where $\mathcal{R}_{\bf k}(t)$ is determined at the time $t_*$:
\[
    \mathcal{R}_{\bf k}=-\biggl[ \frac{H}{\dot
                \phi}\delta\phi_k\biggr]_{t=t_*}.
\]

  Our improvement is associated with calculating the curvature perturbation
  on the basis of exact solutions to the Einstein equations (in the
  zero-order approximation) without using a slow-roll regime in this
  situation. Since we have obtained exact expressions for the cosmological
  parameters at an exit outside the horizon, we need, for confrontation with
  the observational data, to carry out a re-calculation of the cosmological
  parameters to the present epoch. To do that, let us consider the
  post-inflationary evolution of the cosmological perturbations and find
  correction to the cosmological parameters.

  It should be noted that, in the inflationary models under study, there is
  no natural exit from inflation. We therefore suppose that, after scalar
  field decay, the gravitational field turns into Friedmann radiation and
  matter-dominated stages during the same e-folds, characterized by the
  time $t_*$.

  We use the standard method of re-calculating cosmological perturbations
  to Friedmann epochs. It is well known \cite{Riotto02} that, during the
  radiation-dominated stage ($a \sim t^n$, $n = 1/2$), and the
  matter-dominated stage ($a\sim t^{n}$, $n=2/3$), the gravitational
  perturbations (the gravitational potential $\Phi_{k}$) are transformed
  in the following way:

  During the radiation-dominated stage,
\[
    \Phi_{\bf k}=\frac{2}{3}{\mathcal{R}}_{\bf k}.
\]

  During the matter-dominated stage,
\[
    \Phi_{\bf k}=\frac{3}{5}{\mathcal{R}}_{\bf k}.
\]
  We conclude from these relations that the evolution of perturbations
  at superhorizon scales is reduced to simply rescaling their amplitudes.

  Thus the power spectrum of the gravitational perturbations at the
  matter-dominated stage is determined through the power spectrum of
  curvature perturbations:
\[
    {\mathcal{P}}_{\Phi}(MD)=\frac{9}{25} {\mathcal{P}}_{\mathcal{R}}.
\]
  This, in turn, allows us to determine the density contrast and the power
  spectrum of scalar perturbations:
\bear
    \delta_{\bf k}
    \eql  \frac{2}{3}\left(\frac{k}{aH}\right)^2T_g(t,k)\Phi_{\bf k},
\yy
    \mathcal{P}_{\delta}(k,t) \eql
  \frac{4}{25}\left(\frac{k}{aH}\right)^4T_g(t,k)^2\mathcal{P}_{\mathcal{R}},
\ear
  which can be compared with the observational data.

  In the case of arbitrary wavelengths of the perturbations, the transfer
  function can be calculated numerically. The values of the transfer
  function have been tabulated for many cosmological models.

  Thus our improvement introduced at the after-inflationary stage leads
  finally to improved relations containing observational data. Let us apply
  the methods described to some exact solutions.

\section{Cosmological parameters for exact solutions}

\subsection{Power-law inflation.}

  First of all, consider power-law inflation. The scale factor of such a
  model is
\beq
    a(t)=a_s t^m, \cm a_s = \const.
\eeq
  For the spectral indices of scalar and tensor perturbations we obtain
\beq
    n_S-1 = \frac{2}{1-m}=n_G.
\eeq
  If $m>1$ and $|n_S -1|<0.2$, we get a reasonable of the degree of
  expansion: $m>9$.

  The tensor-to-scalar spectral index ratio
\beq
    r=\frac{2}{3-m}
\eeq
  is critically connected with the value $m=3$. When $ m=3$, we have
  $r\to \infty$. For $1\leq m <3$, we have $1\leq r < \infty$, while for
  $m > 3$ we have $-\infty <r<0$. When $m\geq 9$, we obtain $r\geq -1/3$,
  i.e.  $|r|< 0.333 $, and as $m$ grows, the spectral index of tensor
  perturbations becomes smaller and smaller as compared with the scalar
  one. One should note that these results are distinct from those of
  \cite{chnotr05}, which is caused by an error in the expression for the
  spectral index of scalar perturbations, \eq (17) of the cited paper.

\subsection{De Sitter solutions}

  The de Sitter solution describes Universes with and without coordinate
  singularities, with the scale factor
\beq
    a(t)=a_s \sinh (h_*t), \cm a_s = \const,
\eeq
  and
\beq
    a(t)=a_s \cosh (h_*t),
\eeq
  respectively. The spectral indices and their ratio are calculated using
  the general formulas and have the following form:

  For a ``singular'' Universe,
\bearr
    n_S-1=\frac{2[2+\cosh^2(h_*t)]}{\sinh^2(h_*t)},
\nnn
    n_G = -\frac{2}{\sinh^2(h_*t)},\ \ \ r=\frac{2}{5+\cosh^2(h_*t)};
\yyy
  {\mathcal P}_{\mathcal R}=\frac{\coth^4(h_*t)}{8M_p^2h_*[1-\coth^2(h_*t)]},
\nnn
    {\mathcal{P}}_{\mathcal{G}}=\frac{\coth^2(h_*t)}{2\pi^2M_{p}^2},\ \ \
    \frac{T}{S}=4\frac{h_*[1-\coth^2(h_*t)]}{\coth^2(h_*t)}.
\ear

  In the case of the nonsingular de Sitter solution,
\bearr
    n_S-1=\frac{2[2-\sinh^2(h_*t)]}{-\cosh^2(h_*t)},
\nnn
    n_G=\frac{2}{\cosh^2(h_*t)},\ \ \ r=\frac{2}{5-\sinh^2(h_*t)},
\yyy
  {\mathcal P}_{\mathcal R}=\frac{\tanh^4(h_*t)}{8M_p^2 h_*[1-\tanh^2(h_*t)]},
\nnn
   {\mathcal{P}}_{\mathcal{G}}=\frac{\tanh^2(h_*t)}{2\pi^2M_{p}^2},
\nnn
    \frac{T}{S}=4\frac{h_*[1-\tanh^2(h_*t)]}{\tanh^2(h_*t)}
\ear
  The scalar field must be imaginary.

\subsection{Generalized exponential inflation}

  Consider a scale factor which is defined as
\beq
    a=a_{s} e^{(Ae^{\lambda t}+Bt)}.
\eeq
  The Hubble parameter is
\[
    H(t)=A\lambda e^{\lambda t}+B.
\]

  Let us calculate the spectral indices of scalar and tensor modes of
  perturbation as well as the power spectra:
\bearr
      n_{S}-1=\frac{4A \lambda e^{\lambda t}-\lambda(A\lambda e^{\lambda
    t}+B)}{A\lambda^2e^{\lambda t}+(A\lambda e^{\lambda t}+B)^2},
\nnn
    n_{G}=\frac{2A\lambda^2 e^{\lambda t}}{A\lambda^2 e^{\lambda
        t}+(A\lambda e^{\lambda t}+B)^2},
\yyy
    {\mathcal{P}}_{\mathcal{R}}=\frac{(A\lambda e^{\lambda
        t}+B)^4}{8A\lambda^2 M_{p}^2 e^{\lambda t}}, \ \ \
    {\mathcal{P}}_{\mathcal{G}}=\frac{(A\lambda
        e^{\lambda t}+B)^2}{2\pi M_{p}^2}.
\ear

   The ratio of the spectral indices and the tensor-to-scalar ratio at
   crossing of Hubble radius are
\bearr
    r=\frac{A\lambda^2 e^{\lambda t}}{5A\lambda^2 e^{\lambda t}
    +(\lambda e^{\lambda t}+B)^2-\lambda(\lambda e^{\lambda t}+B)},
\nnn
  \frac{T}{S}=4\frac{A\lambda^2 e^{\lambda t}}{(A\lambda e^{\lambda t}+B)^2}.
\ear

\subsection{Exponential-power-law inflation.}

  In this model, the evolution of the scale factor is defined as
\bearr
    a(t)=a_s t^n \exp(h_*t),
\nnn
     n = A^2(2M_P^2), \cm a_s = \const.
\ear
  The cosmological parameters are, in this case, written as
\bearr
    n_S-1=\frac{2(h_*t-n)}{(h_*t+n)^2-n},
\nnn
    n_G=\frac{2n}{n-(h_*t+n)^2}
\yyy
   {\mathcal P}_{\mathcal R}=\frac{t^4}{8M_p^2(h_*t+n)^2[1-t(h_*t+n)]},
\nnn
    {\mathcal{P}}_{\mathcal{G}}=\frac{t^2}{2\pi^2M_{p}^2(h_*t+n)^2}.
\ear
  The ratio of the spectral indices and the tensor-to-scalar ratio are again
  easily calculated:
\bearr
    r=\frac{2n}{(h_*t+n)^2+2h_*t-3n},
\nnn
    \frac{T}{S}=4\frac{t^2}{1-t(h_*t+n)}.
\ear
  Let us find the cosmological parameters for scale factor evolution
\beq
    a = a_{s} e^{\lambda(t-t_{0})^{\mu}}.
\eeq
  By calculations similar to the previous ones, we obtain:
\bearr
    n_{S}-1=\frac{3\mu+1}{\mu(t-t_{0})^{\mu}+\mu-1},
\nnn
    n_{G}=\frac{2}{1+\left[\mu/(\mu-1) \right] (t-t_0)^{\mu} },
\yyy
    {\mathcal{P}}_{\mathcal{R}}=\frac{\lambda^3
        \mu^3}{8M_{p}^2}\frac{(t-t_{0})^{3\mu-2}}{\mu-1},
\nnn
    {\mathcal{P}}_{\mathcal{G}}=\frac{1}{2}\left(
        \frac{\mu\lambda}{M_{p}^2} \right)^2(t-t_{0})^{2\mu-2},
\yyy
    r=\frac{2\lambda t(\mu-1)}{5\lambda\mu(\mu-1)
        +\lambda^2\mu^2t^{\mu+1}-\mu+2},
\nnn
    \frac{T}{S}=\frac{4(\mu-1)}{\mu\lambda(t-t_{0})^{\mu} }
\earn
  In these models, the values of the constants are restricted by $T/S < 0.2$
  according to the observational data \cite{lukash06}.

\section{Discussion}

  Further developing the exact methods of obtaining the key cosmological
  parameters, suggested in \cite{ch04grg, chnotr05}, we have presented exact
  formulas for the power spectra of scalar and tensor (gravitational)
  perturbations and the corresponding spectral indices. (From the formulas
  presented, it is easy to determine the runaway of the spectral indices and
  their form for the models enumerated in this article. These results will
  be given elsewhere.)

  Let us again emphasize the simplicity of the method and its applicability
  to any models in which the gravitational field of the evolving Universe or
  the total energy potential are known (see \cite{ch04grg, chnotr05}).

\Acknow
   {The authors thank the participants of the School-Seminar ``Modern
   Theoretical Problems of Gravitation and Cosmology'' GRACOS-2007,
   especially V. N. Lukash and A. A. Starobinsky, for a discussion of the
   results and a productive debate.}

\end{document}